\documentclass[twoside,fleqn]{article}
\usepackage{espcrc2}
\usepackage{epsfig}
\usepackage{amstex}
\usepackage{amssymb}

\newlength{\hepwidth}
\title{
Decorrelating the topology in full QCD
\hfill\parbox[t][1ex]{\hepwidth}{\normalsize \hfill IFUP-TH 47/96\\
      \hfill hep-lat/9608123}
      }

\author{G. Boyd\address{Dipartimento di Fisica dell'Universit\`a,
    I-56126 Pisa, Italy}
\thanks{
This project was partially supported by the European Union, contract
CHEX--CT92--0051, and by MURST.  GB was supported by the European Union
{\em  Human Capital and Mobility} program under HCM-Fellowship contract
ERBCHBGCT940665. The authors are particularly grateful to Raffaele Tripiccione
for advice and assistance in using the 512 node APE/QUADRICS in Pisa.
}
\\
        with B. All\'{e}s, M. D'Elia, A. Di Giacomo and E. Vicari
       }       

\begin{document}

\begin{abstract}
    We investigate the performance of the hybrid Monte Carlo algorithm
    in
    updating non-trivial global topological structures.  We find that the
    hybrid Monte Carlo algorithm has serious problems decorrelating the global
    topological charge.
    This represents a warning which must be seriously considered when
    simulating full QCD, regardless of the number and type of fermions, with
    this or any similar algorithm. Simulated tempering
    is examined as a means of accelerating the decorrelation.
\end{abstract}

\maketitle

\section{Introduction}
All production runs in full QCD use molecular dynamics based algorithms. For
integer multiples of four staggered or two Wilson fermions one has the exact
hybrid Monte Carlo (HMC), while for other multiples there is the very similar
hybrid molecular dynamics algorithm.

We examine here how effective the HMC is  in decorrelating the topology 
for both full QCD with four staggered fermions and pure $SU(3)$ gauge
theory.  The latter we compare with an over-relaxed heat-bath algorithm.
Although the only HMC algorithm with four staggered fermions has been
investigated, the conclusions drawn are valid for any algorithm based on
molecular dynamics.

An ensemble containing the correct topology\footnote{We regard the topological
  modes to be those which, in the continuum limit, determine the topological
  properties of the lattice.} is crucial for the $\eta^{\prime}$ mass, for
example. It can also become relevant for other quantities, for example the
proton mass, at some level of accuracy. One criterion that the ensemble must
fulfill is that the global topological charge averages to zero; we find that
this is not so for a long run at our smallest quark mass. 

Other simulations involving the topology in the presence of fermions have been
presented in the literature~\cite{Bitar,fermcm,japan}, where the existence of
long range correlations have been reported.

We have a large set of configurations generated with the HMC, originally
prepared with the aim of studying the spin of the proton~\cite{MTCspin}.  As
the operator involves the topology, the longest topological auto-correlation
length determines the number of independent
configurations. This is the auto-correlation of the global topological charge 
on cooled~\cite{coolingmethod} configurations.

The conclusions drawn here require only that our method sees the most slowly
updated modes associated with the `topological' content of the lattice
configurations. For this purpose any reasonable definition of the global
topological charge that is not affected by noise is adequate; we employed the
field theoretic definition of $Q$, measured on cooled configurations.

\section{Results}
We use the $\Phi$ algorithm of reference~\cite{GT187}, with a $16^{3}\times 24$
lattice at a coupling $\beta=5.35$ for full QCD, and a $16^{3}\times 16$
lattice at $\beta=6.00$ for pure SU(3). The two relevant units for comparing
the auto-correlations are then fictitious molecular dynamics time in
units of $\tau$, and wall clock time.

\begin{figure}[tb]
\begin{center}
\leavevmode
\epsfig{file=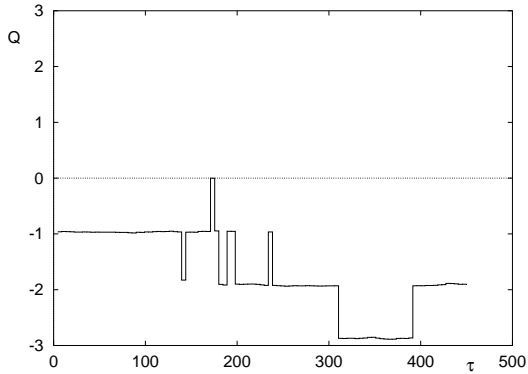,height=5cm}
\caption{
  Time history, in units of molecular dynamics time $\tau$, of the topological
  charge $Q$ for the HMC simulation at $\beta=5.35$ and $m=0.01$ on a
  $16^3\times 24 $ lattice.  }
\label{fig1}
\end{center}
\end{figure}

Let us begin with the results for quark mass $m=0.01$. This yields a pion to
rho mass ratio of $m_{\pi}/m_{\rho}\simeq 0.5$~\cite{MTC92}, so we
are still quite far from the physical value.  The lattice spacing is $a\simeq
0.14$fm (from $m_{\rho}$), which gives a lattice volume of
$V_{3}\simeq(2.2\text{fm})^3$. 

We performed a rather long simulation, with a total length after thermalisation
of $\tau\simeq 500$ in units of fictitious time. Further details may be found
in~\cite{longpaper}. Fig.~\ref{fig1} shows the time history of the topological
charge, where it is clear that the HMC is unable to change the global
topological modes efficiently, leading to very long auto-correlations. The
value of the topological charge got stuck at around $Q\simeq -2$, and its value
averaged over all configurations generated so far is decidedly non-zero:
$\langle Q \rangle = -1.7(4)$.  A very rough estimate of the integrated
auto-correlation time $T_Q$ from a blocking analysis of the data gives
$T_{Q}\gtrsim 2\times 10^2$ in units of molecular dynamics time.  The
simulations were performed on the 25 Gflops APE tower, with around $50\%$
efficiency. On this machine $T_{Q} \simeq 200$ corresponds to about three days
of computer time, a considerable amount.  Notice that most simulations at
comparable values of $\beta$ and $m$ presented in the literature have $\tau
\simeq 100$.

To further investigate the behaviour of the HMC, we have performed simulations
with larger quark masses, $m=0.035$ and $m=0.05$, and in the quenched case,
which represents the large quark mass limit of full QCD. Quenched simulations
were performed at $\beta=6.0$, and here HMC has been compared with the
over-relaxed heat-bath. One cycle of this algorithm consists of 5
microcanonical updates followed by a pseudo-heat-bath update.

\begin{table}[tb]
\begin{small}
\begin{center}
\caption{Results from, and parameters used for the Hybrid Monte Carlo (HMC)
  runs, and, in the last row, the heat-bath over-relaxed (HBOR) run.  The full
  QCD runs with four staggered fermions were performed on a $16^{3}\times 24$
  lattice, the pure SU(3) runs on a $16^{3}\times 16$ lattice. The length of
  each trajectory in fictitious time is $\tau_{\text{traj}}$. The last three
  columns give the integrated auto-correlation time of $Q$ and the plaquette
  $\Pi$ in units of fictitious time and wall clock hours on the 25 Gflops
  Quadrics.}
\label{tab:param}
\vspace{\baselineskip}
\begin{tabular}{cccccc}
\hline\hline
\multicolumn{1}{c}{$\beta$}&
\multicolumn{1}{c}{$m$}&
\multicolumn{1}{c}{$\tau_{\text{traj}}$}&
\multicolumn{1}{c}{$T_{Q}(\tau)$}&
\multicolumn{1}{c}{$T_{Q}({\text{hrs}})$}&
\multicolumn{1}{c}{$T_{\Pi}({\tau})$}\\
\hline\hline
5.35 & 0.010   &  0.3  & $\gtrsim$ 200 & $\gtrsim$ 72 & 1.2 \\ 
5.35 & 0.035   &  0.3  & 15   & 2    &2.5  \\ 
5.35 & 0.05    &  0.3  &  6   & 0.3  &2.2  \\ 
\hline                  
6.00 &$\infty$ &  0.3  &  320 & 0.9  & 1.7 \\ 
6.00 &$\infty$ &  0.6  &  140 & 0.4  & 1.7 \\ 
6.00 &$\infty$ &  0.16 &  312 & 0.67 & 2.8  \\ 
6.00 &$\infty$ &  0.32 &  180 & 0.34 & 1.9 \\ 
6.00 &$\infty$ &  0.64 &   72 & 0.12 & 2.4  \\
6.00 &$\infty$ &  0.96 &   43 & 0.08 & 4.1(27s)  \\
6.00 &$\infty$ &  1.50 &   69 & 0.12 & 4.9  \\
6.00 &$\infty$ &  2.00 &   83 & 0.15 & 5.6  \\
\hline\hline 
6.00 &$\infty$ & \multicolumn{1}{c}{HBOR}  &   5.7 & 0.0014 & 0.9(0.9s)  \\ 
\hline\hline 
\vspace{\baselineskip}
\end{tabular}
\end{center}
\end{small}
\end{table}

In Table~\ref{tab:param} we give the parameters used in our simulations and the
estimates of the integrated auto-correlation time of the topological charge
$T_{Q}$. The estimates should be regarded as a lower limit for $m=0.01$, and
with a possible uncertainty of 10-20\% for all other values.

There are three major results that have emerged from this work:
\begin{enumerate}
\item In HMC simulations the integrated auto-correlation time of the global
  topological charge decreases as the length of the trajectory increases. If
  expressed in terms of computer time, then trajectories longer than one unit
  in molecular dynamics time seem to be slower.
\item For pure gauge theory at $\beta=6.0$ the HMC algorithm is about
  two orders of magnitude slower in decorrelating $Q$ than the local
  over-relaxed algorithm.  For our `best' set of HMC parameters we find that
  the plaquette decorrelates about 30 times more slowly, which agrees
  with the results of~\cite{gupcomp}.  The plaquette auto-correlation time in
  units of $\tau$ is constant, though.
  However, $Q$ decorrelates 60 times slower with
  the HMC than with the heat-bath over-relaxed for our
  `best' parameters.  Furthermore the
  auto-correlation of $Q$ is far more sensitive to the choice of parameters
  than the plaquette.
\item The auto-correlation time $T_{Q}$ rapidly increases with decreasing quark
  mass, in terms of both CPU time and molecular dynamics time.  The
  auto-correlation time for the plaquette, on the other hand, remains similar
  in units of $\tau$.  The time required appears to increase faster than the
  $T_{Q}\approx 1/m^{2.5}$ expected~\cite{gupcomp,bitcomp,SGacc} if one uses
  $1/m^{2}_{\pi}=1/m$ as the relevant physical quantity, a factor $1/m$ from
  the matrix inversion, and another $\approx 0.5$ from the change in step size
  and acceptance rate. Note that in our rather long HMC simulation at $m=0.01$
  ($\tau \simeq 500$) the ensemble is not yet sufficient to determine the
  auto-correlation at all well; we cannot even estimate how long the run should
  have been to get $\langle Q\rangle\approx~0$.
\end{enumerate}

\section{Improvements}
It may also be possible to improve the performance of the HMC algorithm at
small masses via the use of simulated tempering~\cite{ST}.  In simulated
tempering, usually the coupling becomes a dynamical parameter in the
simulation. We are currently investigating promoting the quark mass to a
dynamical variable in the simulation. This helps in decorrelating the
topological charge both because there are more fluctuations when the mass is
large, and because the HMC algorithm itself is faster for larger masses. 

In simulated tempering one has an ordered set of masses, chosen such that 
the action histogram of adjacent masses has some overlap. The mass can then 
change to an adjacent mass if the configuration lies in this overlapping
region, ie., if the configuration is a representative member of the probability
distribution for each of the masses. 

We are currently investigating simulations using masses between 0.01 and 0.035.
From these preliminary investigations we estimate that the improvements
will more than compensate the additional overhead if the masses used range
between $m\lesssim 0.01$ and 0.02.

Finally, we mention that algorithms that are not based on equations of motion,
(for example, L\"uscher's~\cite{luscherferm} multi-boson algorithm), may
perform better with respect to the topological modes than the HMC algorithm.
This is also being examined.

\section{Conclusions}
In conclusion we have shown that the HMC algorithm has serious problems
decorrelating global topogical modes, more serious than those associated with
commonly studied quantities. For full QCD the algorithm appears to slow down 
much more rapidly than previously expected.
We stress again that this warning must be
seriously considered when simulating full QCD.  It is especially important when
studying quantities related to the topology, but may be less relevant in the
calculation of the mass spectrum (with the exception of, for example, the
$\eta^{\prime}$ mass), since an effective decoupling from the topological modes
is expected. Note also that this slowing down is independent of both the number
of fermions and type of fermion simulated, as the underlying dynamics of the
algorithm are the same as the cases studied here.



\end{document}